\documentclass[letter]{aa}

\usepackage[varg]{txfonts}
\usepackage{epsfig,graphicx,natbib,url,twoopt}
\usepackage[varg]{txfonts}
\usepackage{hyperref}          
\usepackage{float}
\hypersetup{
 colorlinks=true,  
 urlcolor=blue,    
 linkcolor=red,     
 citecolor=blue 
}

\def\kms{\hbox{km$\;$s$^{-1}$}}
      
\def\mAA{m\AA}
\def\Halpha{\mbox{H\hspace{0.1ex}$\alpha$}}
\def\Hbeta{\mbox{H\hspace{0.1ex}$\beta$}}

\def\CaK{\ion{Ca}{ii}~K}

\def\MgH{\ion{Mg}{ii}~h}

\def\IRIS{{IRIS}}

\begin{document}

\title{Signatures of ubiquitous magnetic reconnection in the lower solar atmosphere}

\author{Jayant Joshi\inst{1,2} 
\and
Luc H. M. Rouppe van der Voort\inst{1,2} 
\and
Jaime de la Cruz Rodr{\'i}guez\inst{3}
}
\authorrunning{J. Joshi et al.}

\institute{Institute of Theoretical Astrophysics,
  University of Oslo, %
  P.O. Box 1029 Blindern, N-0315 Oslo, Norway
\and
 Rosseland Centre for Solar Physics,
  University of Oslo, %
  P.O. Box 1029 Blindern, N-0315 Oslo, Norway
\and
  Institute for Solar Physics, Dept. of Astronomy, Stockholm University, AlbaNova University Centre, 10691, Stockholm, Sweden
  }


\abstract
{
Ellerman Bomb-like brightenings of the hydrogen Balmer line wings in the quiet Sun (QSEBs) are a signature of the fundamental process of magnetic reconnection at the smallest observable scale in the solar lower atmosphere.
%
We analyze high spatial resolution  observations (0\farcs1) obtained with the Swedish 1-m Solar Telescope to explore signatures of QSEBs in the \Hbeta\ line.
%
We find that QSEBs are ubiquitous and uniformly distributed throughout the quiet Sun, predominantly occurring in intergranular lanes. 
We find up to 120 QSEBs in the FOV for a single moment in time; this is more than an order of magnitude higher than the number of QSEBs found in earlier \Halpha\ observations. 
This suggests that about half a million QSEBs could be present in the lower solar atmosphere at any given time.
The QSEB brightening found in the \Hbeta{} line wings also persist in the line core with a temporal delay and spatial offset towards the nearest solar limb.
Our results suggest that QSEBs emanate through magnetic reconnection along vertically extended current sheets in the solar lower atmosphere.
The apparent omnipresence of small-scale magnetic reconnection may play an important role in the energy balance of the solar chromosphere.    
}


\maketitle

\section{Introduction}
\label{sec:introduction}

Magnetic reconnection is at the root of a wide range of magnetic activity phenomena in the solar atmosphere and is the fundamental driver of, for example, solar flares and coronal mass ejections.
In the deep solar atmosphere, magnetic reconnection is manifested as so-called  Ellerman Bombs (EB): small but intense brightenings of the extended wings of the hydrogen Balmer-$\alpha$ line (\Halpha).
EBs are traditionally found in complex bipolar active regions, particularly during episodes of vigorous flux emergence, and the observation that they are invisible in the \Halpha\ line core locate the site of reconnection below the chromospheric canopy of fibrils
\citep{2011ApJ...736...71W, 
2013ApJ...774...32V}. 

The advanced numerical simulations of 
\citet{2017ApJ...839...22H, 
2019A&A...626A..33H} 
and \citet{2017A&A...601A.122D} 
have reinforced the interpretation of EBs as markers of small-scale photospheric magnetic reconnection. 
EBs are further characterized as a transient phenomenon with lifetimes on the order of  minutes and a morphology in slanted-view \Halpha-wing images that appears like flames: bright, upright, 1--2 Mm tall and rapidly flickering on timescales of seconds 
\citep{2011ApJ...736...71W}. 
We refer to 
\citet{2013JPhCS.440a2007R} 
for a review of EB properties and the introduction of 
\citet{2019A&A...626A...4V} 
for an overview of EB visibility in different diagnostics.

\citet{2016A&A...592A.100R} 
recently discovered a phenomenon of Ellerman Bomb-like brightenings in the quiet Sun, far away from regions with strong magnetic activity. 
%
These quiet-Sun Ellerman Bombs (QSEBs) are smaller in size and exhibit weaker enhancement of their \Halpha{} wings than their active region counterparts
\citep{2016A&A...592A.100R, 
2017ApJ...845...16N}. 
QSEBs are also thought to originate through magnetic reconnection in the photosphere
\citep{2016A&A...592A.100R, 
2018MNRAS.479.3274S}, 
similar to EBs. 

Here, we further advance the observational characterization of QSEBs by analysis of high resolution observations in the Balmer-$\beta$ (\Hbeta) line.
The shorter wavelength of the \Hbeta\ line allows higher spatial resolution and enhanced intensity contrast compared to \Halpha\ observations.
The latter stems from the wavelength dependence of the Planck function which results in higher contrast at shorter wavelengths for the enhanced temperature in the reconnection site.
We find much higher numbers of QSEBs compared to the earlier \Halpha\ observations.
This apparent ubiquity of QSEBs raises interest in the impact of low atmospheric reconnection events onto the total energy budget of the solar atmosphere. 
Further, we see evidence for dynamical progression of the reconnection process from lower to higher altitude.
These findings highlight the potential of high resolution observations in probing the smallest observable magnetic reconnection events in the solar atmosphere.

%
%
%
%
%
  
\begin{figure*}[!ht]
\centering
\includegraphics[width=0.95\textwidth]{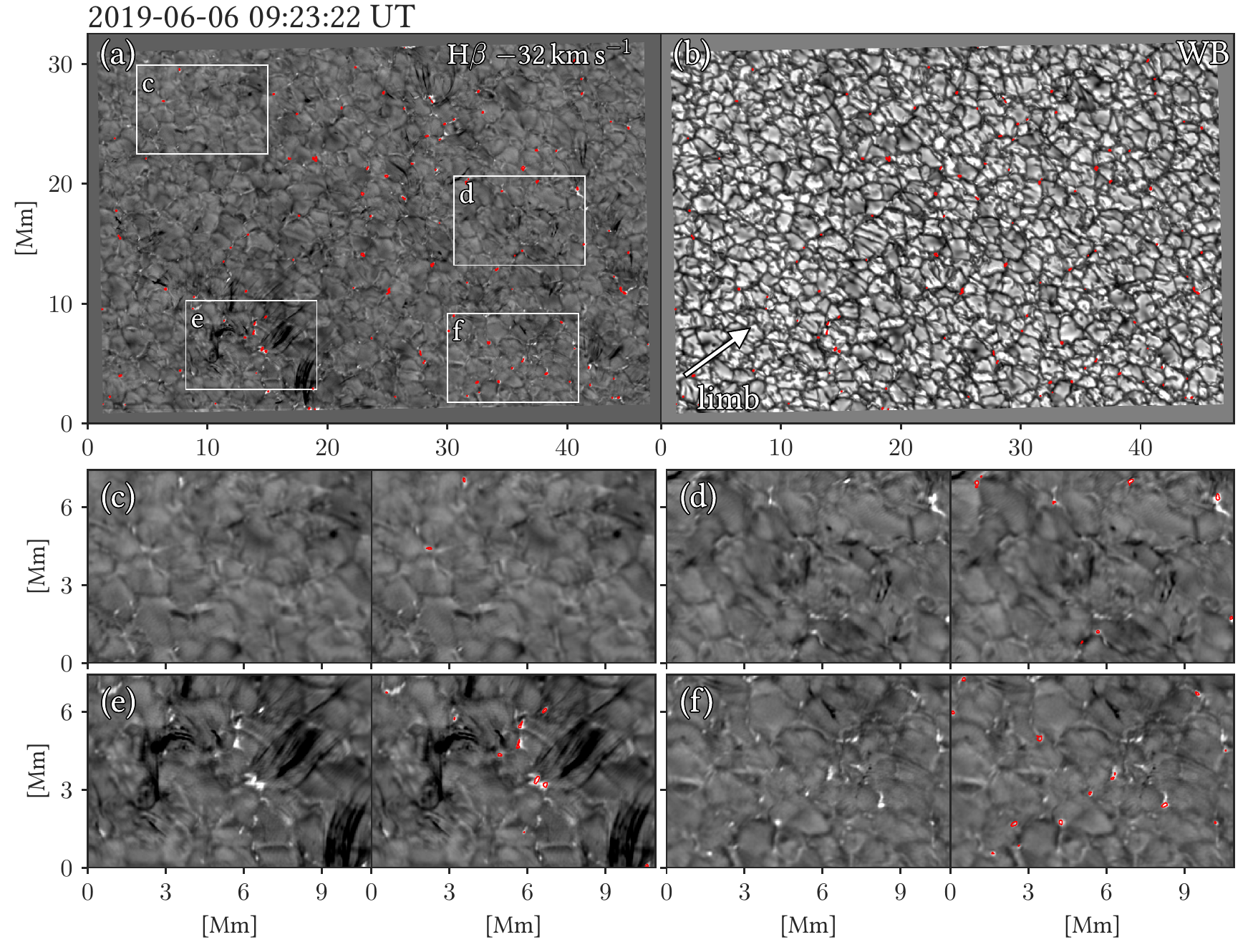}
\caption{\label{fig:overview}%
Ubiquitous QSEB detections using the \textit{k}-means clustering technique.
(a): the observed FOV in the \Hbeta\ blue wing. 
The red contours indicate the location of QSEBs. 
(b): the co-temporal continuum WB image. 
(c--f): zoom of the four different areas indicated by the white boxes in (a); each panel shows duplicate images, one with the red contours indicating the location of QSEBs and one without the contours. The arrow in panel~(b) marks the direction of the closest solar limb. 
An animation of this figure is available at: \url{https://folk.uio.no/jayantj/QSEB/movie1.mp4}.
}
\end{figure*}

\begin{figure*}[!ht]
\centering
\includegraphics[width=0.95\textwidth]{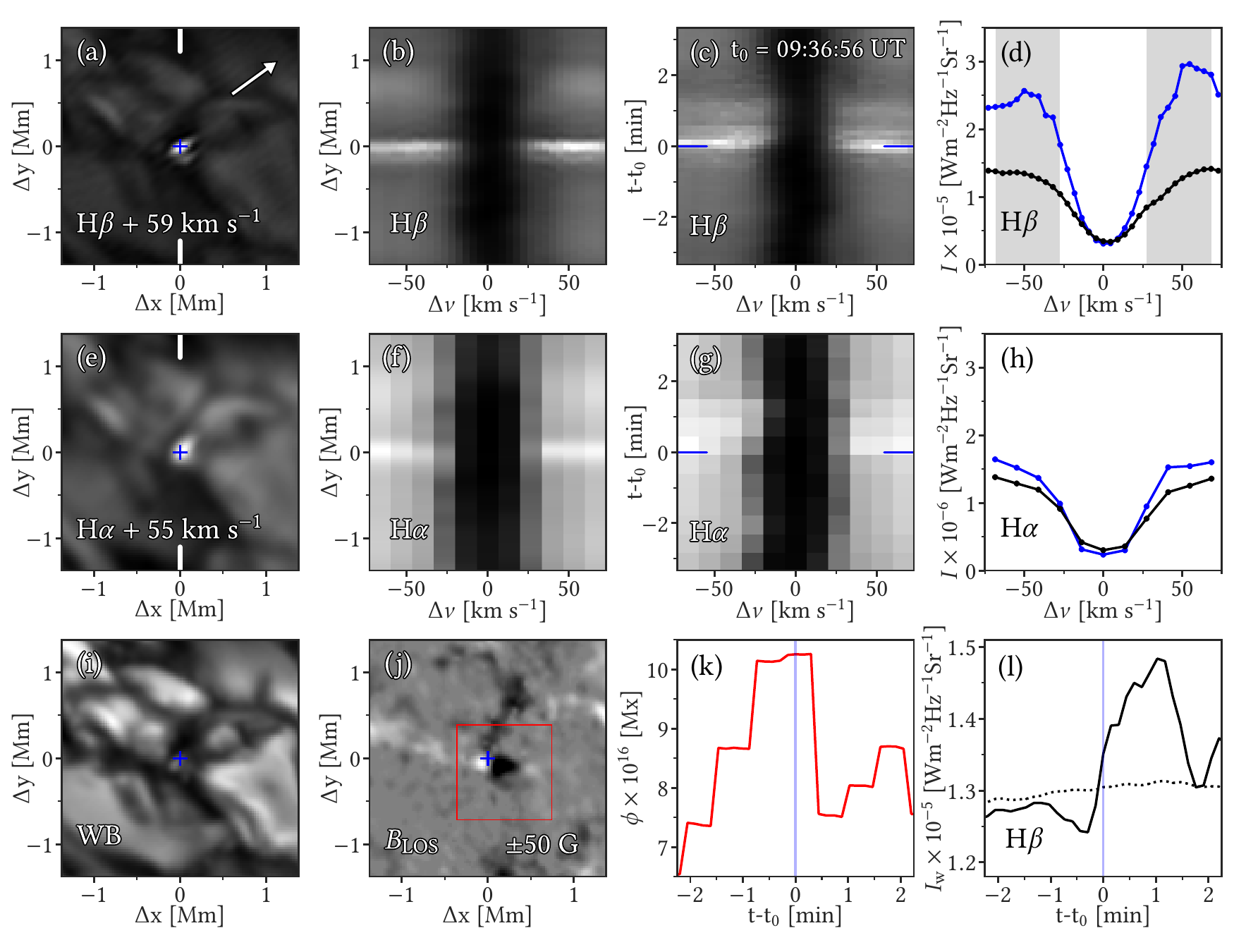}
\caption{\label{fig:qseb_zoom_1}%
Details of a QSEB in \Hbeta{} and \Halpha.
(a): the QSEB observed in the \Hbeta{} red wing. (b): spatial variations in the \Hbeta{} line profile along the y-axis crossing the QSEB and marked by the two vertical white lines in (a). Here the spectral dimension is presented in terms of Doppler offset. (c): temporal variations in the \Hbeta{} line profile for a location in the QSEB indicated by the blue plus sign in (a). (d): the \Hbeta{} line profile at the brightest pixel in the QSEB and an average quiet-Sun reference profile (black line). (e--h): similar to (a--d) but in \Halpha. (i): the corresponding WB image and (j): map of $B_{\rm{LOS}}$. (k): the evolution of the $B_{\rm{LOS}}$ flux ($\phi$) within the red box shown in (j). (l): light curves of the intensity variations of the QSEB (solid) and averaged over the FOV presented in this figure (dotted). Both the curves are average of intensities in the \Hbeta{} line wings ($I_{\rm{w}}$) as marked by the grey shaded area in (d). The arrow in (a) shows the direction to the closest solar limb. An animation showing evolution of the QSEB is available at: \url{https://folk.uio.no/jayantj/QSEB/movie2.mp4}.
}
\end{figure*}

\begin{figure*}[!ht]
\centering
\includegraphics[width=0.95\textwidth]{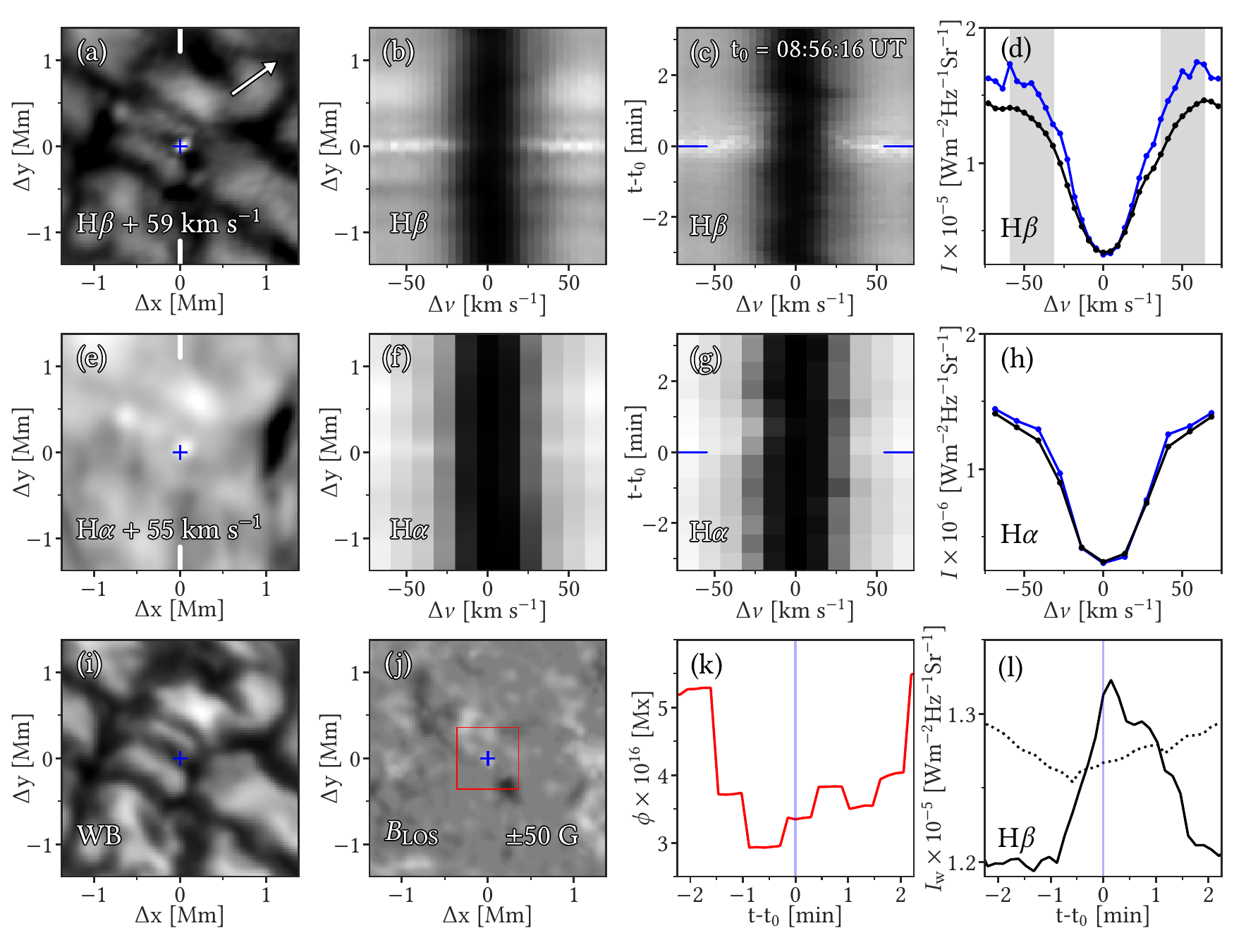}
\caption{\label{fig:qseb_zoom_2}%
Example QSEB which is small, short-lived and less bright compared to the QSEB example in Fig.~\ref{fig:qseb_zoom_1}. This figure has the same format as Fig.~\ref{fig:qseb_zoom_1}. An animation of this figure is available at: \url{https://folk.uio.no/jayantj/QSEB/movie3.mp4}.
}
\end{figure*}

\begin{figure*}[!ht]
\centering
\includegraphics[width=0.95\textwidth]{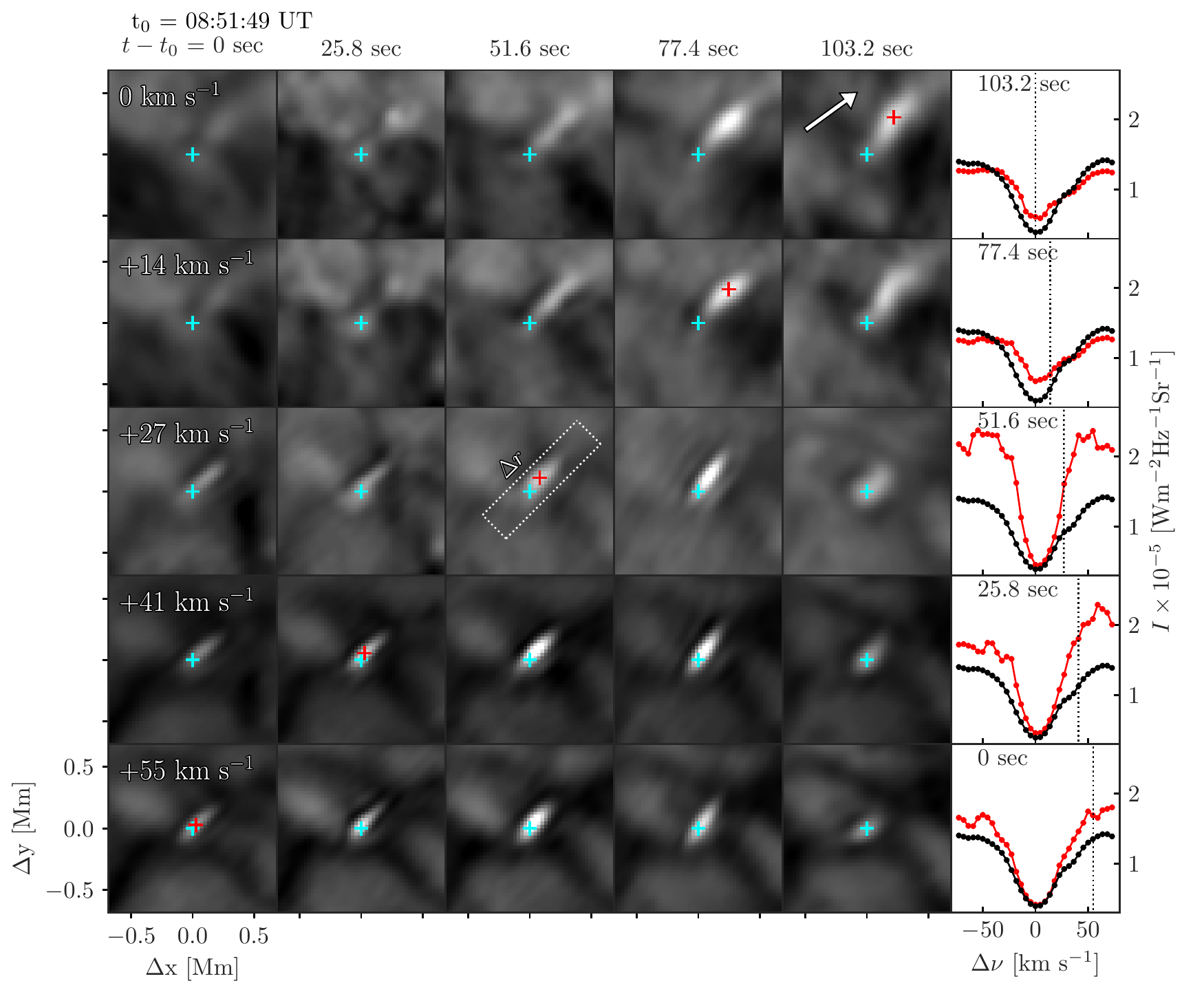}
\caption{\label{fig:lcb}%
Temporal evolution of a QSEB. 
Time is progressing along the rows of \Hbeta\ images from left to right, Doppler offset is varying along the columns from line core at the top to far wing in the bottom.
The rightmost column shows the evolution of the QSEB \Hbeta{} line profiles in red (the location of the line profile is marked with the red plus sign in the images), the black line is a reference profile averaged over the presented FOV. 
The vertical dotted line marks the Doppler offset in the corresponding row. 
The \Hbeta\ profile is selected from the location of maximum intensity at that Doppler offset within the area of the QSEB.
The cyan plus signs mark the centre of the FOV. 
Images in a particular row are displayed on the same intensity scale.
The dotted rectangle on the central image shows the area used to create the space-time map displayed in Fig.~\ref{fig:spacetime}(b).          
}
\end{figure*}

\begin{figure}[!ht]
\centering
\includegraphics[width=0.40\textwidth]{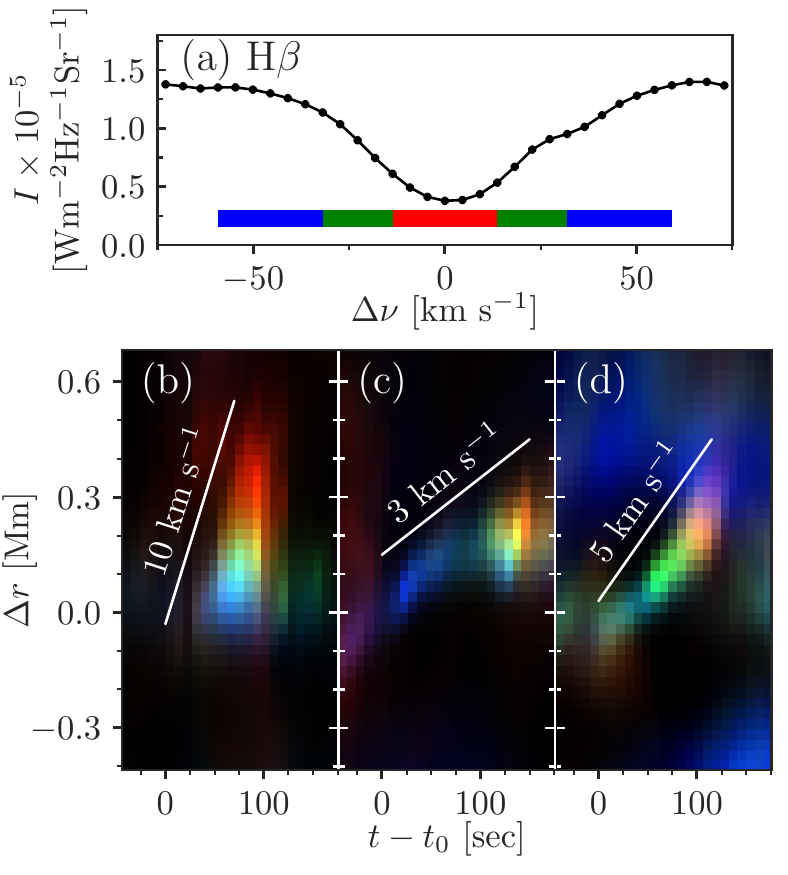}
\caption{\label{fig:spacetime}%
Space-time maps demonstrating progression of the QSEB brightening from the \Hbeta{} line-wings to the line-core in time and space. 
The space-time maps in panels (b--d) are constructed as RGB color images with the \Hbeta{} line core image in the red, flanks in the green, and wings in the blue channel.
The space-time maps in panels~(b), (c), and (d) corresponds to the QSEBs presented in Fig.~\ref{fig:lcb}, \ref{fig:lcb_B1}, and \ref{fig:lcb_B2}, respectively. The displayed space-time maps are created by averaging intensities along the width of the dotted rectangle plotted on the central image of the respective figures and $\Delta r$ representing the length of the rectangles. The white lines in panels~(b--d) indicate the apparent speed by which the QSEB brightening progresses from the wings to the core in the \Hbeta{} line.    
}
\end{figure}
%

\section{Observations and data processing}
\label{sec:obs}

The observations were obtained with the CHROMIS 
and CRISP
\citep{2008ApJ...689L..69S} 
instruments at the Swedish 1-m Solar Telescope 
\citep[SST, ][]{2003SPIE.4853..341S} 
on 6 June 2019.
The target was a Quiet Sun region at $(x,y)=(611\arcsec,7\arcsec)$ and the time series has a duration of 1~h. 
CHROMIS sampled the \Hbeta\ spectral line at 4861~\AA{} at 35 line positions between $\pm$1.371~\AA{} with 74~\mAA\, steps between $\pm$1.184~\AA{}. 
The temporal cadence was 8.6~s.
CHROMIS has a transmission profile FWHM of 100~\mAA{} at 4860~\AA{} and a pixel scale of 0\farcs038.
The CHROMIS instrument has an auxiliary wide-band (WB) channel that is equipped with a continuum filter centered at 4845~\AA\ (FWHM=6.5~\AA).

CRISP ran a programme sampling the \Halpha, \ion{Fe}{i}~6173~\AA{}, and \ion{Ca}{ii}~8542~\AA{} spectral lines at a cadence of 35.9~s.
For this study, we concentrated on the \Halpha\ observations (sampled at 11 line positions between $\pm$1.5~\AA{} with 30~\mAA{} steps) and maps of the line of sight component of the magnetic field ($B_\textrm{LOS}$) derived from Milne-Eddington inversions of the spectropolarimetric \ion{Fe}{i}~6173~\AA{} observations (sampled at 13 line positions with $\pm$32~\mAA{} steps plus continuum at $+680$~\mAA).

High spatial resolution was achieved by the combination of good seeing conditions, the adaptive optics system and the high-quality CRISP and CHROMIS reimaging systems 
\citep{2019A&A...626A..55S}. 
We further applied image restoration using the multi-object multi-frame blind deconvolution 
\citep[MOMFBD, ][]{2005SoPh..228..191V} 
method. 
The data was processed with the CRISP and CHROMIS data processing pipelines 
\citep{2015A&A...573A..40D, 
2018arXiv180403030L}. 

We have performed Milne-Eddington inversions of the \ion{Fe}{i}~6173~\AA\ line data to infer the line-of-sight component of the magnetic field utilizing a parallel C++/Python implementation\footnote{\url{https://github.com/jaimedelacruz/pyMilne}} \citep{2019A&A...631A.153D}.
\section{Data Analysis}
\label{sec:analysis}

\subsection{Identification of QSEBs}
\label{sec:identity}
Signatures of EBs and QSEBs are readily observable in the \Halpha{} 
and \Hbeta{} 
lines; the wings of these lines show enhancement whereas the line core remains unaffected.
To identify this telltale spectral signature of QSEBs, we used the \textit{k}-means clustering technique 
\citep{everitt_1972}, 
which is widely used for the characterization of a variety of solar phenomena and observations.
Examples include the classification of \MgH{} and k line profiles observed with \IRIS{} 
\citep{2019ApJ...875L..18S}, 
the identification of \MgH{} and k spectra in flares 
\citep{2018ApJ...861...62P}, 
and \CaK{} observations of on-disk spicules 
\citep{2019A&A...631L...5B}. 
Using \textit{k}-means clustering, we classified the \Hbeta{} observations into a hundred different line profiles.
Out of the hundred \Hbeta{} line profiles, we identified profiles that show the signatures of QSEBs, i.e., the enhanced wings and unaffected line-core.
%
The red contours indicate QSEB detections in Fig.~\ref{fig:overview}, which shows an example \Hbeta\ blue wing image alongside with the co-temporal continuum WB image.
%
%
We will delineate the details of the \textit{k}-means classification methodology to identify QSEBs in a follow-up paper.

Figure~\ref{fig:overview} shows that the QSEBs are almost uniformly distributed throughout the FOV.
From the zoom-in on four different areas, 
presented in Fig.~\ref{fig:overview}, 
it is evident that QSEBs are predominantly located in the inter-granular lanes.
We found a total of $\sim$2800 events in 420 CHROMIS spectral scans within the FOV of 60\arcsec$\times$40\arcsec.
The number of detected events varies from scan to scan, 
clearly correlated
to variations in the seeing conditions.
In the best quality scans we found up to 120 QSEBs in the FOV suggesting that there could be as many as half a million QSEBs present on the solar surface at any given time.  

\subsection{Selected examples of QSEBs}
\label{sec:qseb}
QSEBs vary significantly in their properties, such as lifetime, brightness, and size. 
%
In this study we present a few specific examples of QSEBs that are representative of all the QSEBs and demonstrate the variations in different properties. 

The QSEB displayed in Fig.~\ref{fig:qseb_zoom_1} is in the category of brighter, bigger, and longer lived 
QSEBs. 
In the \Hbeta{} red wing intensity map, the QSEB appears as a bright flame-like structure elongated in the direction of the closest solar limb. 
The \Hbeta{} spectra of the QSEB show the typical enhancement of the line wings while the line core remains unchanged compared to the average background spectrum.
%
The QSEB is also visible in the intensity map obtained in the \Halpha\ red wing. However, the \Halpha{} spectra exhibit very little enhancement in the line-wings. 
The WB image 
suggests that the QSEB is located in an intergranular lane and we note the absence of any photospheric magnetic bright point (BP).   
The QSEB is located at the intersection of two magnetic patches with opposite polarities.
The animation of Fig.~\ref{fig:qseb_zoom_1} shows that the opposite polarity patches merge, leading to magnetic flux cancellation. 
Panels (k) and (l) illustrate that at the onset of the QSEB, the unsigned $B_{\rm{LOS}}$ flux ($\phi$) starts decreasing. 
Moreover, the light-curve of the QSEB  in panel~(l) indicates that it has a lifetime of around two minutes.
 
Figure~\ref{fig:qseb_zoom_2} presents a QSEB event that is smaller, less bright, and shorter lived compared to the QSEB in Fig~\ref{fig:qseb_zoom_1}.
The \Hbeta{} spectra show weak enhancement of the line wings which is found only in six pixels which indicates a QSEB area of around 8000~km$^2$. 
On the other hand, the QSEB presented in Fig.~\ref{fig:qseb_zoom_1} is approximately five times bigger. 
In the \Halpha{} observations the QSEB is hardly distinguishable from the background.    
The QSEB is among the shortest lived QSEBs with a lifetime of $\sim$35 seconds which is evident from the light-curve of the QSEB and average background intensity plotted in Fig.~\ref{fig:qseb_zoom_2}(l). 
Furthermore, the temporal evolution  of the \Hbeta{} line profile in the QSEB shows the enhancement above the averaged background profile only in four time steps; see the animation of Fig.~\ref{fig:qseb_zoom_2}. 
This QSEB also appears in an intergranular lane and has very weak magnetic field strength ($B_{\rm{LOS}}$ < 10 G).  

Two more QSEB examples are shown in Appendix~\ref{app:examples_qseb} in the same format as Fig.~\ref{fig:qseb_zoom_1} and \ref{fig:qseb_zoom_2}. 
In terms of properties, the QSEB event displayed in Fig.~\ref{fig:qseb_zoom_A1} is very similar to the QSEB in Fig.~\ref{fig:qseb_zoom_1}. 
The QSEB is located in intergranular lanes harbouring multiple $B_{\rm{LOS}}$ patches with mixed polarities and the evolution of unsigned $B_{\rm{LOS}}$ flux suggests a co-temporal flux cancellation episode.   

Finally, the QSEB shown in Fig.~\ref{fig:qseb_zoom_A2} is an example of a QSEB that occurs on top of a photospheric magnetic BP (see panels~(a) and (i)).
Our dataset has ample examples of QSEBs appearing co-spatially with photospheric magnetic BPs.
A detailed analysis of lifetime, area and brightness distribution of the QSEBs along with an analysis of the magnetic field topology 
will be presented in a followup paper.

\subsection{\Hbeta{} line-core brightening associated with QSEBs}
\label{sec:lcb}
As mentioned before, the typical characteristics of EBs and QSEBs are the enhanced \Halpha\ wings and unaltered line core. 
The \Hbeta{} observations presented in this letter 
show overwhelming evidence that the QSEB's brightening also persists in the \Hbeta{} line core.
In this section and Appendix~\ref{app:examples_lcb}, we provide a few examples of the \Hbeta{} line-core brightening associated with QSEBs. 

Figure~\ref{fig:lcb} shows the temporal evolution of a QSEB at different line positions in the \Hbeta{} line. 
At first, the QSEB appears in the line-wings (at Doppler offsets of 41--55~\kms{}), while it is barely visible in the intensity maps of the line flanks (14--27~\kms{}) and absent in the line core. 
As the evolution of the QSEB progresses, the brightening appears in the line flanks and finally in the line core, approximately 50-–80 s after the QSEB onset. 
By the time the brightening begins to appear in the line core, it weakens in the wings.
Apart from its gradual progression to the line core from the wings, the QSEB also shows a systematic spatial displacement between the line wings and line core. 
The QSEB brightening in the line core is located more towards limb-ward direction as compared to the wings, this indicates temperature enhancement at higher altitude. 

Two additional examples of the QSEB brightening in the \Hbeta{} line-core are demonstrated in Figs.~\ref{fig:lcb_B1} and ~\ref{fig:lcb_B2} of Appendix~\ref{app:examples_lcb}, which present a qualitatively similar scenario as described above.  

We show space-time maps in Fig.~\ref{fig:spacetime} for the QSEBs displayed in Figs.~\ref{fig:lcb}, ~\ref{fig:lcb_B1}, and ~\ref{fig:lcb_B2}, which provide clear illustration of the gradual shift in the QSEB's brightening from the \Hbeta\ line wings towards the line core as well as the gradual spatial displacement. 
From the diagonal slope in the space-time maps we estimate that the QSEB's brightening moves from the line wings towards the line core with apparent velocities ranging between 3-10~\kms{}. 

\section{Discussions and conclusions}
\label{sec:conlusions}

We present \Hbeta\ observations that reveal the ubiquitous presence of QSEBs in quiet Sun.
In the highest quality images, recorded during the best seeing conditions, we find up to 120 QSEBs in uniform distribution over the FOV.
This number of QSEBs is more than an order of magnitude higher than in the \Halpha\ observations of
\citet{2016A&A...592A.100R} 
who found less than 10 QSEBs in their best images. 
We estimate that about half a million QSEBs could be present in the lower solar atmosphere at any given time. 
%
The QSEBs mainly arise in the intergranular lanes, but also occasionally appear co-spatially with photospheric magnetic BPs.
We find significant variation in QSEB properties: ranging from larger ($\sim$40,000~km$^2$) and brighter QSEBs that last for several minutes to smaller ($~\sim$8,000~km$^2$) and relatively less bright QSEBs that have a lifetime of a few tens of seconds.
%
Comparing co-temporal \Halpha\ and \Hbeta\ observations, we conclude that the latter diagnostic is more suited to study QSEBs. 
%
%
QSEBs show more pronounced intensity enhancement of the \Hbeta\ wings and small and weak QSEBs are difficult to distinguish from the background in the \Halpha{} observations.
We attribute the unambiguous identification of QSEBs in \Hbeta{} observations to higher spatial resolution that can be achieved at shorter wavelengths. 
%
%
Another major factor is the intrinsic higher intensity contrast at shorter wavelengths where the Planck function has larger variation for different temperatures. 
%
%
The wings of hydrogen Balmer lines form in the lower parts of the solar atmosphere and are more coupled to the local thermodynamic conditions.
%
%
Therefore, a temperature increase due to magnetic reconnection will result in a higher intensity enhancement of the \Hbeta{} line wings compared to the \Halpha{} line. 

By definition, the quiet-Sun is where the Sun exhibits minimal solar activity, and no large-scale magnetic flux emergence occurs. 
On the other hand, new magnetic flux does appear 
in the interior of supergranular cells
\citep[see,][ and references therein]{2011SoPh..269...13T, 
2016ApJ...820...35G, 
2017ApJS..229...17S};
\citet{2017ApJS..229...17S} found the flux emergence rate of 1100 Mx cm$^{-2}$ day$^{-1}$. 
Moreover, with time, 
a significant fraction of newly appeared magnetic elements dissipate by cancellation with opposite polarity patches
\citep{2016ApJ...820...35G,
2017ApJS..229...17S}.
So, flux cancellation prevails throughout the quiet-Sun photosphere and might be responsible for QSEBs.
%
We found episodes of flux cancellation in the photosphere that are spatially and temporally coinciding with QSEBs. This supports the idea that magnetic reconnection in the photosphere is the fundamental mechanism for QSEBs \citep{2016A&A...592A.100R, 
2018MNRAS.479.3274S}. 
Nonetheless, not all the QSEBs show flux cancellation in our data and many appear with unipolar $B_{\rm{LOS}}$ patches.
%
%
We note that lower spatial and temporal resolution in our $B_{\rm{LOS}}$ maps as compared to the \Hbeta{} observations make it difficult to detect weak and small opposite polarity patches.
Furthermore, projection effects in our off-center observations may hide magnetic patches rooted in deep intergranular lanes. 

The three-dimensional (3D) radiation magnetohydrodynamic (MHD) simulations by
\citet{2017ApJ...839...22H, 
2019A&A...626A..33H}  
successfully reproduce observable properties of EBs and suggest that they originate through magnetic reconnection in vertical or nearly vertical elongated current sheets in the lower solar atmosphere.
In particular, 
\citet{2019A&A...626A..33H} 
explained the co-spatial and temporal presence of EBs and UV bursts 
\citep{2014Sci...346C.315P} 
observed by 
\citet{2015ApJ...812...11V, 
2016ApJ...824...96T, 
2020A&A...633A..58O}. 
Their simulations show that the reconnection in the upper part of the current sheet (in the chromosphere/transition region) produces enhanced \ion{Si}{iv} emission associated with UV bursts, whereas EBs with enhanced Balmer wings occur in the lower part of the current sheet, in the photosphere.
Indications of a spatial offset between EBs and UV bursts in off-center observations were first found by
\citet{2015ApJ...812...11V} 
and later clearly demonstrated by
\citet{2019ApJ...875L..30C}. 
Our observations of a flame-like morphology of QSEBs aligned along the limbward direction as well as associated brightenings in the \Hbeta{} line core, which appears with a spatial offset towards the limb, indicate that QSEBs also originate in vertically elongated current sheets stretching between the photosphere and chromosphere.    
%
Moreover, the temporal delay in the \Hbeta{} line core brightening compared to the wings, suggest progression of the magnetic reconnection from the photosphere towards the chromosphere with a speed of 3-10~\kms.

In the flux emergence simulations by
\citet{2017ApJ...839...22H, 
2019A&A...626A..33H}, 
the magnetic reconnection responsible for EBs takes place in a $\cup$-shape magnetic field topology, where the two opposite sides get advected through a converging photospheric flow;
a scenario earlier put forward by e.g.,  
\citet{2008ApJ...684..736W}. 
On the contrary, 
\citet{2017A&A...601A.122D} 
found the magnetic reconnection in $\Omega$-shape topology producing EBs in her 3D MHD simulations. 
Given the ubiquity of QSEBs and the minimal magnetic activity in quiet-Sun, it is less likely that QSEBs also originate in the magnetic field topology described above which usually features in regions of flux emergence. 
Therefore we propose an alternative scenario for QSEBs: mixed polarity magnetic patches in intergranular lanes produced by the local dynamo action
\citep{1993A&A...274..543P, 
1999ApJ...515L..39C, 
2007A&A...465L..43V} 
might generate small-scale current sheets and consequently, magnetic reconnection.
However, the proposed mechanism requires scrutiny through 3D numerical simulations.

A series of recent numerical studies 
\citep{2018ApJ...862L..24P, 
2019ApJ...872...32S, 
2020ApJ...891...52S} 
suggest that ubiquitous photospheric flux cancellation driven by magnetic reconnection could act as a mechanism for chromospheric and coronal heating.
%
%
Here we present the first-ever unambiguous detection of ubiquitous magnetic reconnection in the lower solar atmosphere. In terms of ubiquity, this reminds one of the nanoflare mechanism of 
\citet{1988ApJ...330..474P}, 
a scenario of numerous and continuous small-scale reconnection events in the upper atmosphere that can explain coronal heating. 
%
%
It will be of great interest to quantify the energy budget of QSEBs and explore their role in chromospheric and coronal heating. 
%

\begin{acknowledgements}
The Swedish 1-m Solar Telescope is operated on the island of La Palma
by the Institute for Solar Physics of Stockholm University in the
Spanish Observatorio del Roque de los Muchachos of the Instituto de
Astrof{\'\i}sica de Canarias.
The Institute for Solar Physics is supported by a grant for research infrastructures of national importance from the Swedish Research Council (registration number 2017-00625).
This research is supported by the Research Council of Norway, project number 250810, and through its Centres of Excellence scheme, project number 262622.
This study benefited from discussions during the workshop ``Studying magnetic-field-regulated heating in the solar chromosphere'' (team 399) at the International Space Science Institute (ISSI) in Switzerland.
We made much use of NASA's Astrophysics Data System Bibliographic Services.
JdlCR is supported by grants from the Swedish Research Council
(2015-03994), the Swedish National Space Agency (128/15). This project has received funding from the European Research Council (ERC) under the European Union's Horizon 2020 research and innovation programme (SUNMAG, grant agreement 759548).
\end{acknowledgements}

\bibliographystyle{aa-note}
\bibliography{qseb} 

\begin{appendix}
\section{Additional QSEB examples}
\label{app:examples_qseb}
Two additional examples of QSEB are presented here, for discussion see Sect~\ref{sec:qseb}.

\begin{figure*}[hbt!]
\centering
\includegraphics[width=0.90\textwidth]{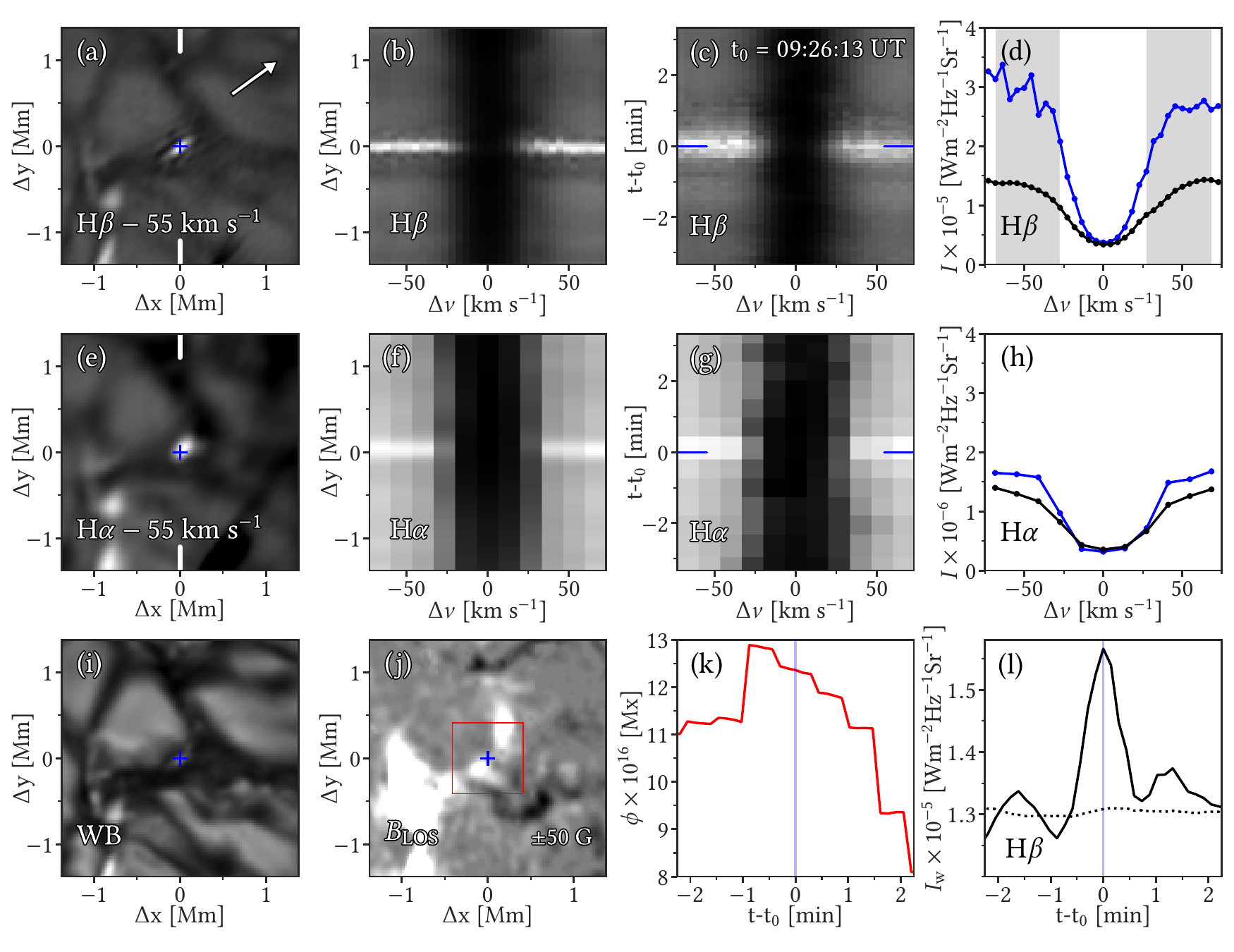}
\caption{\label{fig:qseb_zoom_A1}%
An additional example of QSEB in the same format as Fig.~\ref{fig:qseb_zoom_1}. An animation of this figure is available at the following link: \url{https://folk.uio.no/jayantj/QSEB/movieA1.mp4}.
}
\end{figure*}

\begin{figure*}[hbt!]
\centering
\includegraphics[width=0.95\textwidth]{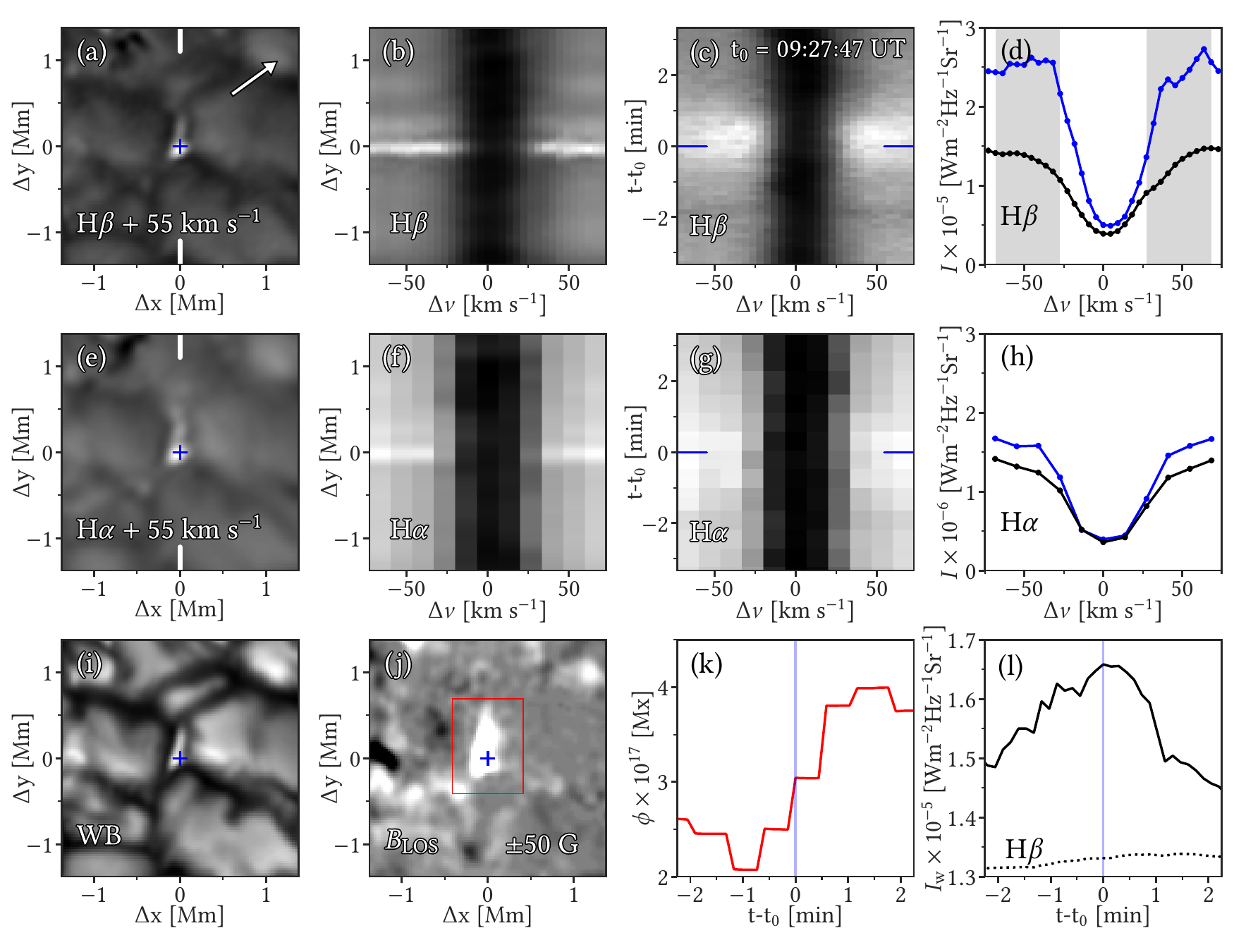}
\caption{\label{fig:qseb_zoom_A2}%
An example of QSEB that appears co-spatially with a photospheric magnetic bright point. This figure is in the same format as Fig.~\ref{fig:qseb_zoom_1}. An animation of this figure is available at the following link: \url{https://folk.uio.no/jayantj/QSEB/movieA2.mp4}.
}
\end{figure*}

\section{Additional examples of the \Hbeta{} line-core brightening}
\begin{figure*}[hb!]
\centering
\includegraphics[width=0.95\textwidth]{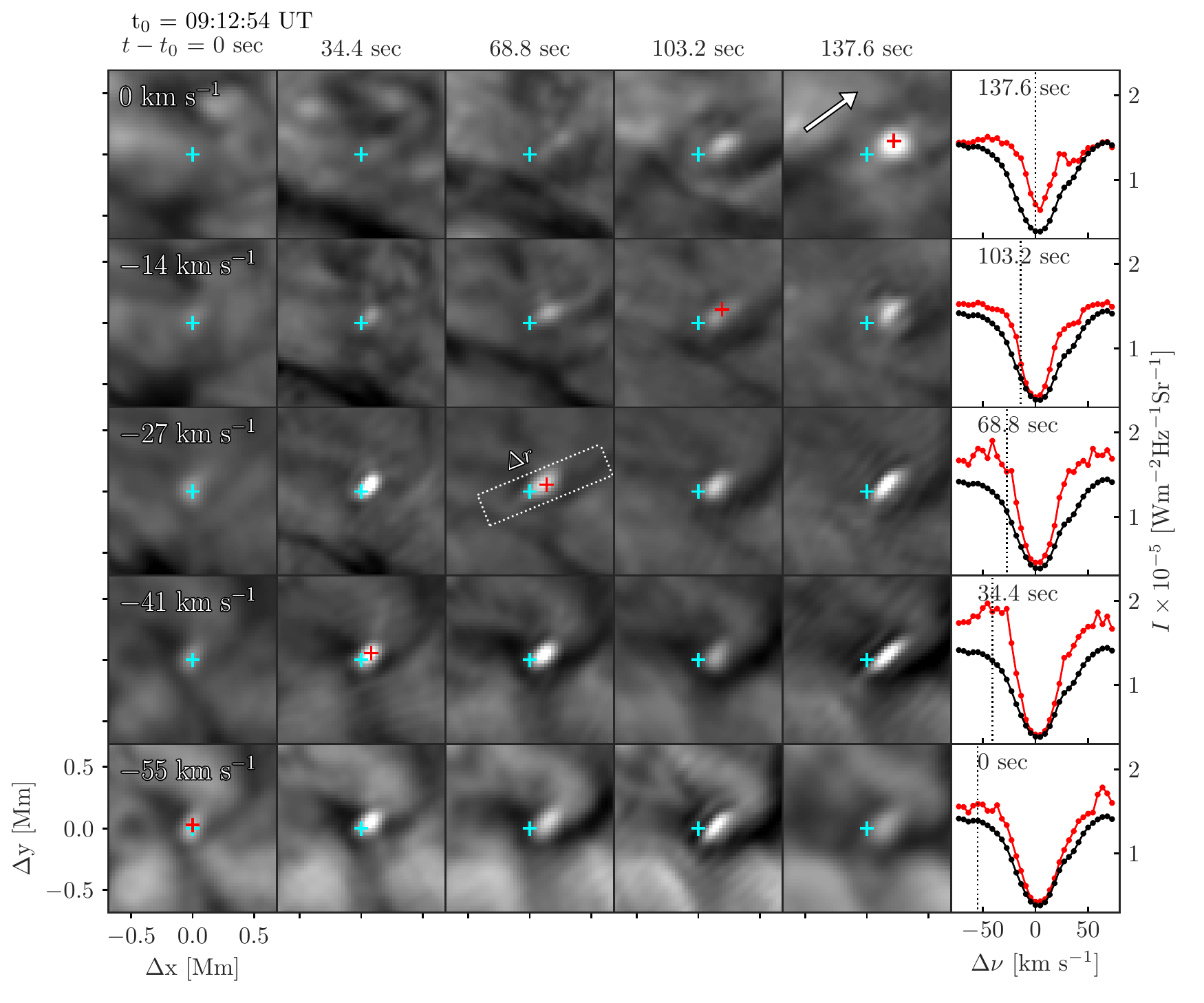}
\caption{\label{fig:lcb_B1}%
An additional example of QSEB that exhibits brightening in the \Hbeta{} line-core. This figure is in the same format as Fig.~\ref{fig:lcb}.
}
\end{figure*}

\begin{figure*}[hbt!]
\centering
\includegraphics[width=0.95\textwidth]{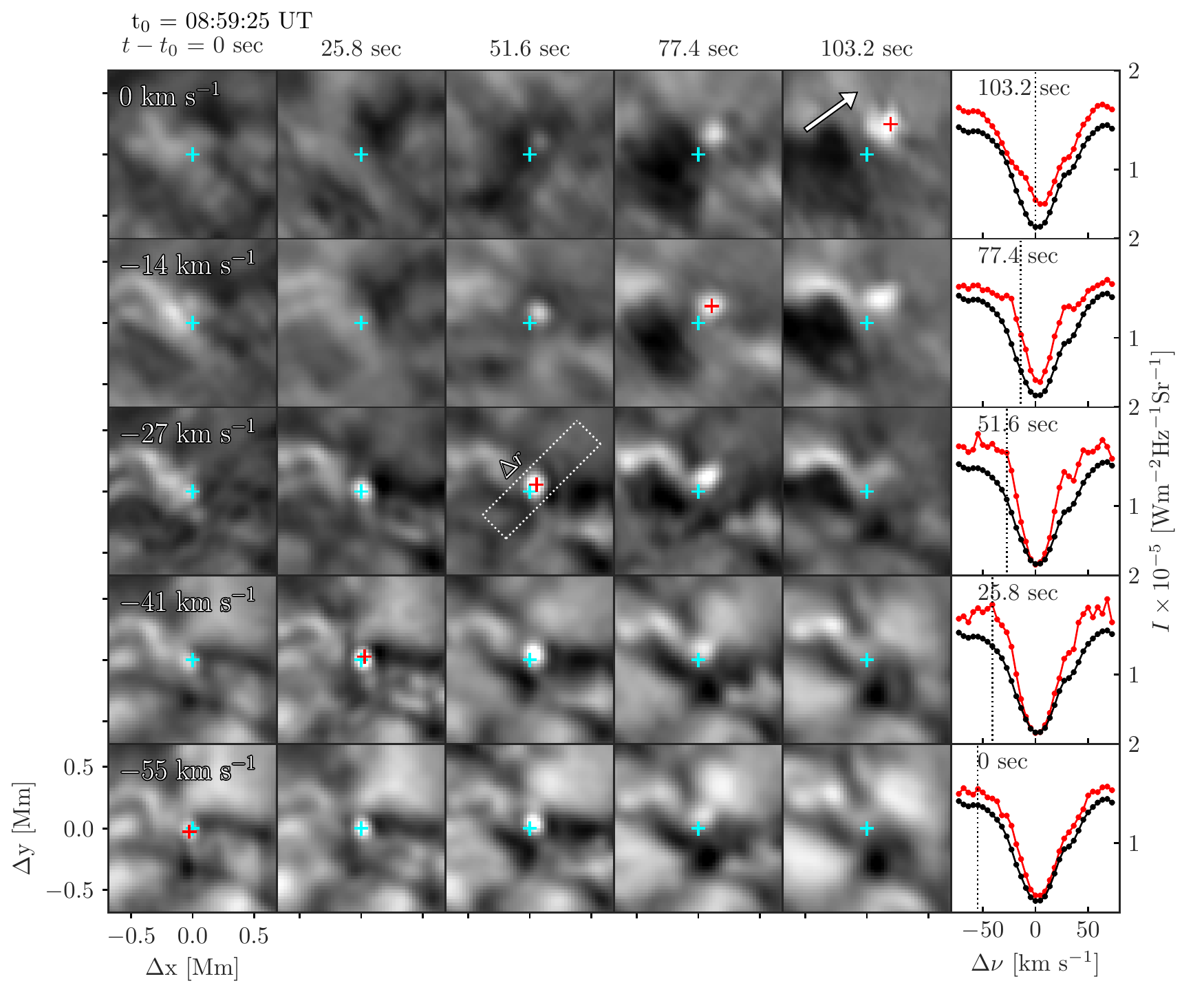}
\caption{\label{fig:lcb_B2}%
An example of QSEB which is smaller and less brighter compared to the QSEBs shown in Fig.~\ref{fig:lcb} and ~\ref{fig:lcb_B1}; and exhibits brightening in the \Hbeta{} line-core. This figure is in the same format as Fig.~\ref{fig:lcb}.}
\end{figure*}

\begin{figure*}[hbt!]
\centering
\includegraphics[width=0.95\textwidth]{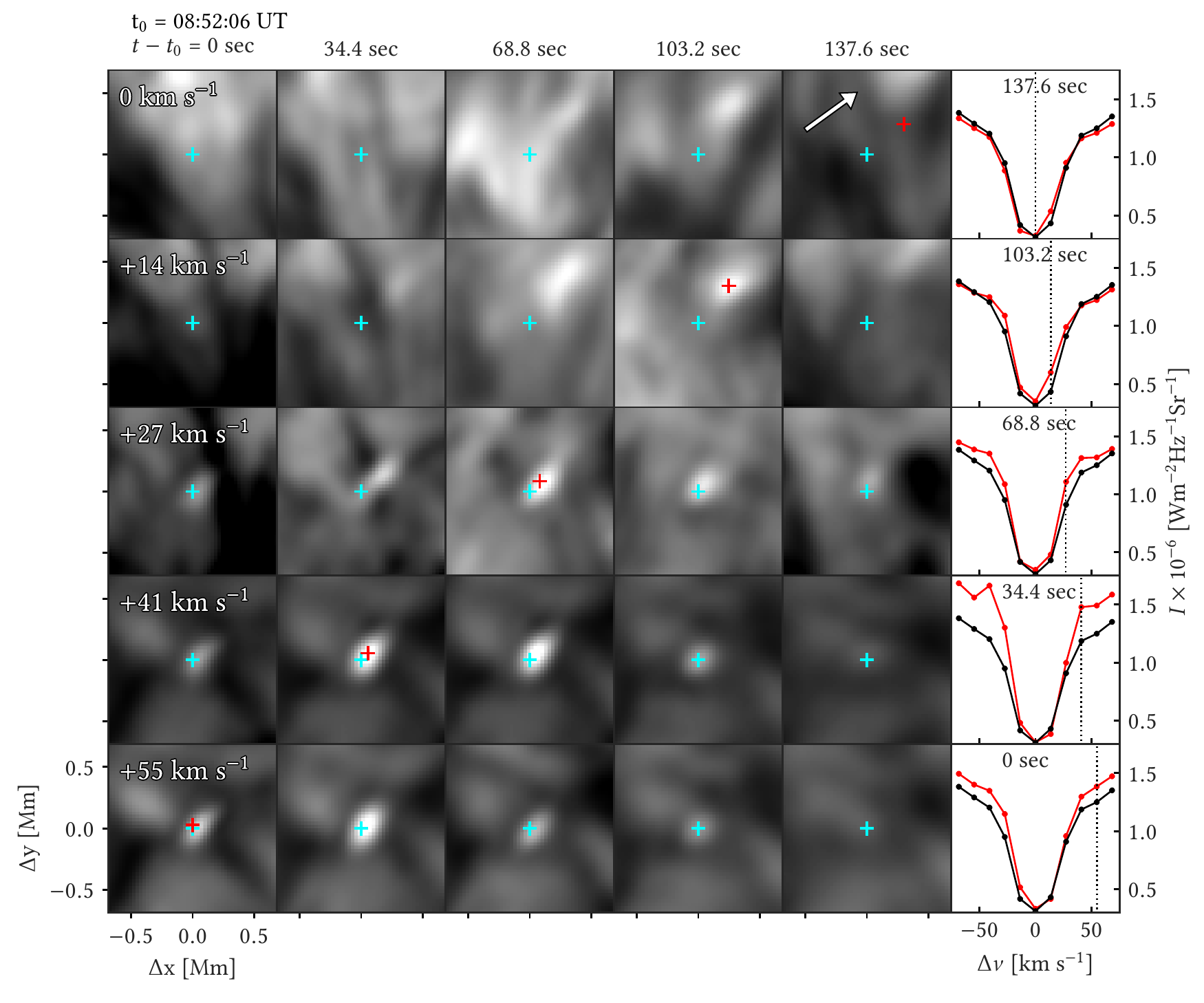}
\caption{\label{fig:lcb_halpha}%
Shows the same QSEB which is presented in Fig.~\ref{fig:lcb}, but illustrates the \Halpha{} observations instead of \Hbeta{}.
}
\end{figure*}

Here, we show the temporal evolution of two QSEBs in the \Hbeta{} line in Fig.~\ref{fig:lcb_B1}, and ~\ref{fig:lcb_B2}, in a similar manner as Fig.~\ref{fig:lcb}.
Figure~\ref{fig:lcb_halpha} illustrates the temporal evolution of the same QSEB presented in Fig. 4, but in the \Halpha{} line instead of \Hbeta{}. 
Figure~\ref{fig:lcb_halpha} suggests that the QSEB also shows the progression from the wings to the flank in the \Halpha{} line. 
However, the QSEBs do not appear in the \Halpha{} line core, unlike the \Hbeta{} observations. 
Signature of the QSEB in \Hbeta{} line core and its absence in the \Halpha{} line core suggest a difference in line core opacity of these two lines.    

\label{app:examples_lcb}
%
%

\end{appendix}

\end{document}